\documentclass{ifacconf}

\usepackage{graphicx}      
\usepackage{natbib}        
\usepackage[dvipsnames]{xcolor}
\usepackage{amsfonts}
\usepackage{amsmath}
\usepackage{amssymb}
\usepackage{mathtools}
\usepackage{multirow}

\usepackage{array}
\newcolumntype{P}[1]{>{\centering\arraybackslash}p{#1}}
\newcolumntype{M}[1]{>{\centering\arraybackslash}m{#1}}

\begin{document}
\begin{frontmatter}

\title{PID Tuning using Cross-Entropy Deep Learning: a Lyapunov Stability Analysis}
\author[First]{Hector Kohler}
\author[Second,fourth]{Benoit Clement}
\author[First,Second]{Thomas Chaffre}
\author[First]{Gilles Le Chenadec}

\address[First]{Lab-STICC UMR CNRS 6285, ENSTA Bretagne, Brest, France (e-mail: name.surname@ensta-bretagne.fr).}
\address[Second]{Centre for Maritime Engineering, Flinders University, Australia\\(e-mail: name.surname@flinders.edu.au).}
\address[fourth]{CROSSING IRL CNRS 2010, Adelaide, Australia.}

\begin{abstract}                
Underwater Unmanned Vehicles (UUVs) have to constantly compensate for the external disturbing forces acting on their body. Adaptive Control theory is commonly used there to grant the control law some flexibility in its response to process variation. Today, learning-based (LB) adaptive methods are leading the field where model-based control structures are combined with deep model-free learning algorithms. This work proposes experiments and metrics to empirically study the stability of such a controller. We perform this stability analysis on a LB adaptive control system whose adaptive parameters are determined using a Cross-Entropy Deep Learning method.

\end{abstract}

\begin{keyword}
Underwater Vehicle, Adaptive Control, Deep Learning, Lyapunov Stability.
\end{keyword}

\end{frontmatter}
\section{Introduction} \label{intro}

Operating in a constantly disturbed environment, Unmanned Underwater Vehicles (UUVs) must compensate for wave and current-induced forces acting on their body. In this context, a common practice is to exploit a Proportional-Integral-Differential (PID) control law, whose parameters are obtained generally using optimal model-based control theory and are then kept constant during operation. 

However, their performances decay under intensive process variation and uncertainties such as those found in the underwater environment, where the vehicle state measurements are limited and the uncertainty on the external disturbance is high. The performance of PID regulators can be improved by accommodating such changes using online tuning techniques, such as adaptive control theory \cite{ODwyer2003HandbookOP,CAMS21a,chaffre_learning-based_2021}.

Adaptive control methods are widely used in the context of dynamical systems and provide what seems to be an ideal framework for automatic tuning of regulators \cite{strm1989AdaptiveC,Parks1981StabilityAC}. New tuning techniques are emerging from the development of data-driven theory \cite{Brunton2019DataDrivenSA},

with Genetic algorithm also suggested to tune PID regulators \cite{Jayachitra2014GeneticAB}. Deep Reinforcement Learning techniques (DRL) in particular, have shown great performance as optimization methods. They exploit the strong representation abilities offered by artificial neural networks (ANNs) to build nonlinear mapping functions between sensor feedback and control inputs/controller parameters, directly in closed-loop systems. Such methods are denoted as learning-based (LB). Despite their increased performance, their usage in UUV applications, where process observability is low, is still limited due to the stability analysis being hardly practicable to conduct when using ANNs.

In this work, we propose an adaptive LB controller where the PID gains are automatically adjusted by an ANN for the task of minimizing setpoint tracking error (i.e. position and orientation) of UUVs under Lyapunov stability constraints. The ANN is trained with a Cross-Entropy Method (CEM) \cite{Rubinstein1997OptimizationOC} which, unlike gradient-based optimization methods, is a direct policy search technique. Moreover, the considered application allows us to apply the classic Lyapunov stability analysis.

This paper is organized as follows: Section \ref{related-works} presents some related usage of Deep Learning methods for the control of UUVs. In Section \ref{auv}, we present the considered robotic platform with its modeling and the simulation framework we used to train our controller. Our strategy for stability analysis is presented in Section \ref{methodology}. In Section \ref{lb-tuning} we describe our learning-based (LB) PID tuning approach. A complete description of our usage of Deep Learning is provided in Section \ref{cross-entropy} with the complete hyperparameters choice. The experiments are described in Section \ref{exp-res} with an analysis of the results, leading to some open questions.

\section{Related work} \label{related-works}
Numerous aspects of PI-PID tuning are discussed in \cite{ODwyer2003HandbookOP}. The most common method consists of auto-tuning where an experiment is performed in open or closed-loop to estimate the process model by using recursive least squares. The idea is that if a second-order model can be generated, it can then be used to make optimal pole-placement. However, the underwater context does not allow such procedures as open-loop experiments are too dangerous for the vehicles and the limited observability of the system/environment does not allow satisfying system identification.
Recently, Deep Learning methods have been successively used in the field of PID tuning.
Among the various existing techniques, Deep Reinforcement Learning (DRL) \cite{Sutton-Barto2ed} has shown great performance in the adaptive control of UUVs \cite{Wang2018ReinforcementLA}, \cite{Knudsen2019DeepLF}. These methods are used to iteratively build an estimate of an optimal policy function, mapping the process states to the control parameters.

The DRL-based methods oblige to frame the control process as a Markov Decision Process (MDP). The associated Bellman's equations are then estimated using ANNs, usually involving Gradient-based learning. In contrast, we proposed in this paper to use CEM to optimize the ANN, removing the need for the MDP formulation.

\section{AUV modeling and simulation} \label{auv}

\subsection{AUV model} \label{auv-model}
A mathematical model of a ROV platform can be derived using the general equations of motion for a marine craft, which can be written in the vectorial form according to \cite{fossen2011} as:

\begin{equation}
\left\{
\begin{aligned}
&\dot{\eta}=J_\Theta(\eta)\nu \\
&M\dot{\nu}+C(\nu)\nu+D(\nu)\nu+g(\eta)=\delta+\delta_{cable}
\end{aligned}
\right.
\end{equation}

where $\eta$ and $\nu$ are the position and velocity vectors respectively, $\delta$ is the control force vector and $\delta_{cable}$ is the vector describing the umbilical forces from the cable attached to the ROV. The RexROV2 is an ROV-type platform provided in UUV Simulator \cite{Manhaes_2016}. It is propelled by 6 thrusters (complete details on its equation of motions are provided in \cite{Berg2012DevelopmentAC}, \cite{McCue2016HandbookOM}, \cite{YCMLW15}. The control vector $u$ is obtained by transforming the actuator force vector: $\delta=T(\alpha)Ku$,

where $T(\alpha)\in\mathbb{R}^{n\times r}$ is the thrust allocation matrix; $K$ is the thrust coefficient matrix; $\delta$ is the control force vector in $n$ degrees of freedom (DoF) and $u\in\mathbb{R}^r$ is the actuator input vector. The package \cite{Manhaes_2016} allows us to design a vector of thruster contribution for every DoFs.

\subsection{ROS packages} \label{ros-pkg}
We used the ROS-based package called \textit{UUV Simulator} proposed in \cite{Manhaes_2016}, to train our controller.

This package provides a Gazebo-based simulation of underwater environments and the possibility to use the existing RexROV2 platform described in \cite{Berg2012DevelopmentAC} as a test vehicle. In addition, it is also possible to simulate several disturbances including sea currents. They are simulated as a uniform force in the Gazebo world and represented by a linear velocity $v_c$ (in $m.s^{-1}$), a horizontal $h_c$, and a vertical angle $j_c$ (in radians).

\section{Our learning-based adaptive control framework} \label{methodology}

In this paper, we address the problem of setpoint regulation where the UUV's objective is to minimize its tracking error in all 6 DoFs. The tracking error $e$ is defined as the error between, the current UUV's position and orientation, and a fixed desired position and orientation as:
\begin{align}
    e = \eta_d - \eta \label{track_err}.
\end{align}
The position and orientation of the UUV is:
\begin{align}
    \eta = (x, y, z, roll, pitch, yaw) \label{uuv_state},
\end{align}
and the desired setpoint: 
\begin{align}
    \eta_d = (0, 0, 0, 0, 0, 0).
\end{align}

In our LB adaptive control framework, a parametrization such as an artificial neural network, of the PID's gains is learned (in simulation or in the real world). In the LB control framework's testing phase, at every state measurement, a control step is performed. In our case, during a control step, the state of the control process is used as input of a trained neural network. The outputs of the network parametrize the PID gains (see section \ref{lb-tuning}) used to compute the control input: the UUV's thrusters. The PID control law is defined as the following:
\begin{align}\label{PID}
u=B^{-1}[\ J^{T}(\eta)\big(k_p e+k_i \int_{0}^{t} e(\tau) d \tau-k_d \dot{\eta}\big)+g(\eta)\ ],
\end{align}
where $u\in \mathbb{R}^6$ are the control inputs;
the PID gain matrices $k_p, k_i, k_d \in \mathbb{R}^{6 \times 6}$ are the output of a neural network; $\eta\in\mathbb{R}^6$ is the current UUV's state, i.e. position and orientation (Eq. \ref{uuv_state}); $\dot{\eta}\in\mathbb{R}^6$ is the temporal derivative of the state vector;

$g(\eta)$ is the sum of external forces acting on the UUV's body (in our case, gravity); $e\in\mathbb{R}^6$ is the tracking error (Eq. \ref{track_err}); $B^{-1}$ is the fixed known thrusters allocation matrix mapping the control input $u$ into a combination of thruster power inputs resulting to the desired movement and $J^{T}(\eta)$ is a transformation matrix.

The principal advantage of the considered PID regulator (Eq. \ref{PID}) is that global stability and convergence analysis of the system has been well formalized by Fossen \cite{fossen2011}. Lyapunov stability theory \cite{LyapunovFunc} is straightforward to apply with such a control law. Following \cite{fossen2011}, there exists a Lyapunov function for the considered UUV such that:
\begin{align} \label{Lyapunov-fct}
V(x)&=\frac{1}{2} x^{T}\left[\begin{array}{ccc}
M_{\eta}^{-1} & \alpha I & 0 \\
\alpha I & k_{p} & k_{i} \\
0 & K_{i} & \alpha k_{i}
\end{array}\right] x ,
\end{align}
where $\alpha\in\mathbb{R}$ is a small positive constant and the PID gains are $k_p, k_i, k_d \in \mathbb{R}^{6 \times 6}$; $M_{\eta}^{-1}$ is related to the UUV's mass and can be computed from the current $\eta$; $x$ is the control loop's state (not to be mistaken with the UUV's state): $x=\big[p, \eta, \int_{0}^{t} e(\tau) d \tau\big]^{T} \in \mathbb{R}^{18}$ and $p = M_{\eta} \dot{\eta}^T \in \mathbb{R}^6$ is the generalized momentum depending on the UUV's mass and velocity. 

The control loop is stable at state $x$ if:
\begin{align} \label{stab_1}
V(x) > 0 \text{ and } \dot{V}(x) < 0.
\end{align}

Following Lyapunov stability theory \cite{LyapunovFunc}, there exist theoretical constraints on the  gain matrices $k_p, k_i, k_d$ and the small constant $\alpha $ such that local stability is guaranteed when the initial conditions of the systems are closed to $\boldsymbol{x}=0$. According to the proposed Lyapunov function (Eq. \ref{Lyapunov-fct}), the vehicle stability and convergence to steady-state are guaranteed (Eq. \ref{stab_1}) if the following constraints are satisfied:\\
\begin{equation} \label{constraints}
\hspace{-1px}\left\{
\begin{array}{l}
k_{d} >M_{\eta}, \\
k_{i} >0, \\
k_{p} >k_{d}+\frac{2}{\alpha} k_{i}, \\
\frac{1}{2}(1-\alpha) k_{d} -\alpha M_{\eta}+\frac{\alpha}{2} \sum_{i=1}^{6}\left(\eta_{i}-\eta_{i d}\right) \frac{\partial M_{\eta}}{\partial \eta_{i}} >0,\\
\alpha > 0,
\end{array}
\right.
\end{equation}

We found that, when limiting the parameter space to value satisfying (Eq. \ref{constraints}), the resulting space is so small that the benefits of adaptive control are merely preserved. Therefore, we propose not to take into account the stability constraints in the parameters optimization. Our objective is then to assess to what extent the resulting solution, obtained from an ANN, can still hold some stability components.

In order to facilitate the stability analysis, we aim at reducing the dimension of the space depicted in Eq.\eqref{constraints}. First, we can transform\eqref{constraints} into equalities as follows:
\begin{equation} \label{sol}
\left\{
\begin{array}{l}
k_{d} = M_{\eta} + M_1, \\
k_{i} = 0 + M_2, \\
k_{p} = k_{d}+\frac{2}{\alpha} k_{i} + M_3, \\
\alpha = \| \frac{-k_d}{( -  k_{d} -2 M_{\eta}+ \sum_{i=1}^{6}\left(\eta_{i}-\eta_{i d}\right) \frac{\partial M_{\eta}}{\partial \eta_{i}})} \|_{max} + \epsilon,
\end{array}
\right.
\end{equation}

where $M_1$, $M_2$ and $M_3$ are three $6\times6$ positive matrices and $\epsilon$ is a small positive constant. With this transformation \eqref{sol}, the fulfillment of the Lyapunov stability constraints \eqref{constraints} can now be verified by only assessing the value of $[M_1, M_2, M_3, \epsilon]\in\mathbb{R}^{3\times6\times6+1,>0}$. In order to further reduce this dimension space, we apply a diagonal transformation on the matrices $M_i$: $M_i=P\Lambda_iP^{-1}$,

where $\Lambda_i\in\mathbb{R}^6$ are positive vectors and $P\in\mathbb{R}^{6\times6}$ is a positive invertible matrix chosen randomly beforehand. Thanks to this transformation, we can now assess the value of $[M_1,M_2,M_3,\epsilon]\in\mathbb{R}^{3\times6\times6+1,>0}$ (and $K_p,K_i,K_d$ with Eq.\ref{sol}) by only accessing:
\begin{equation} \label{param-space}
[\Lambda_1, \Lambda_2,\Lambda_3, \epsilon] \in \mathbb{R}^{19,>0}.
\end{equation}

\section{PID tuning using Deep learning} \label{lb-tuning}
In order to take into account the process uncertainties, we propose to learn a stochastic predictive model $\pi_\omega$ (parameterized by $\omega$) that maps the system state vector $x$ into the controller parameters of the PID law:\\
\begin{equation} \label{desired-policy}
\left\{
\begin{aligned}
&\pi:  \hspace{.2cm} \Omega_x \subset \mathbb{R}^{18} &\mapsto &\hspace{.2cm}\Theta \subset \mathbb{R}^{109}\\
& x=\left[p, \eta, \int_{0}^{t} e(\tau) d \tau\right]^{T} & \mapsto& \hspace{.2cm} \left[k_p,k_i,k_d,\alpha\right] 
\end{aligned}
\right.
\end{equation}

This mapping is composed of two stages. The first stage is a trainable neural network whose input is the state vector $x$, aiming at predicting the vector $[\Lambda_1, \Lambda_2,\Lambda_3, \epsilon] \in \mathbb{R}^{19}$. To this end, we model each variable $n \in \{1,\dots,19\}$ of this vector as a random variable with an independent Normal distribution as:
\begin{equation} \label{action-space}
\mathcal{N}_n(\mu_n, \sigma^2_n)=(2\pi\sigma^2_n)^{-1/2}\exp \{-\frac{1}{2\sigma^2_n}(x-\mu_n)^2\},
\end{equation}
The goal of this network is to output the 19 pairs of $(\mu,\sigma)$ representing the action distributions.
 
We have empirically set up an architecture composed of 2 hidden layers of 32 hidden nodes each, and with the Sigmoid activation function applied to each layer. This results in a total of $(18 + 1) \times 32 + (32 + 1) \times 32 + 2 \times ((32 + 1)\times 19) = 2918 $ parameters ($\omega$) to learn from data. 
In a second stage of the mapping $\pi$, the PID parameters $[\Lambda_1, \Lambda_2,\Lambda_3, \epsilon]$ are obtained by sampling from the resulting Gaussian distributions $\mathcal{N}(\mu,\sigma)$. The Eq. \ref{sol} allows to compute the final PID parameters and using Eq. \ref{PID} the PID control inputs are derived. With this setting, we can more easily assess the Lyapunov stability of the system. The overall control strategy is illustrated in Figure \ref{fig:control_loop}.

\section{Learning with the Cross entropy method} \label{cross-entropy}

We propose to learn/optimize the weights of the neural network with the CEM. The CEM is a direct search method (i.e no gradient is computed). It is an Estimation of Distribution Algorithm (EDA) inspired by Natural Evolution Strategies \cite{Wierstra2008NaturalES}. With CEM, during one iteration $k$, $N$ sets of weights $S^k_{i = 1 ... N}$ are sampled from a Normal distribution directly in the space of weights $R^{2918}$. At each iteration $k$, $N$ evaluations are made to determine the current best weights $S^k_{best = 1 ... N\times \rho}$ with respect to a given cost function. In our case, we used the following classic control performance (based on multi steps within an episode and connected to the tracking error (Eq.\ref{track_err}) index as a cost function to minimize:
\begin{equation}\label{cost-fct}
J = \sum_{steps} \frac{1}{6}\sum^6_{i=1} (\eta_{d,i} - \eta_{i})^2.
\end{equation}
The mean and covariance of the Normal distribution are then updated as the mean and the covariance of the $N \times \rho$ best sets of weights obtained at iteration $k-1$. Noise $\sigma^2_{noise}$ is added to the covariance of the best weights to avoid local optima. We randomly sample the next iteration weights $S^k = \mathcal{N}(\text{mean}(S^{k-1}_{best}), \text{Cov}(S^{k-1}_{best})+\sigma^2_{noise})$. 

\section{Experiments and Analysis} \label{exp-res}
\subsection{Experimental settings}
\subsubsection{Simulation:}
We control a simulated UUV in Gazebo as described in Section \ref{auv}. The simulation is running in real-time and the frequency of the control loop is $20$ Hz.

\begin{figure}
    \centering
    \includegraphics[width=0.9\linewidth]{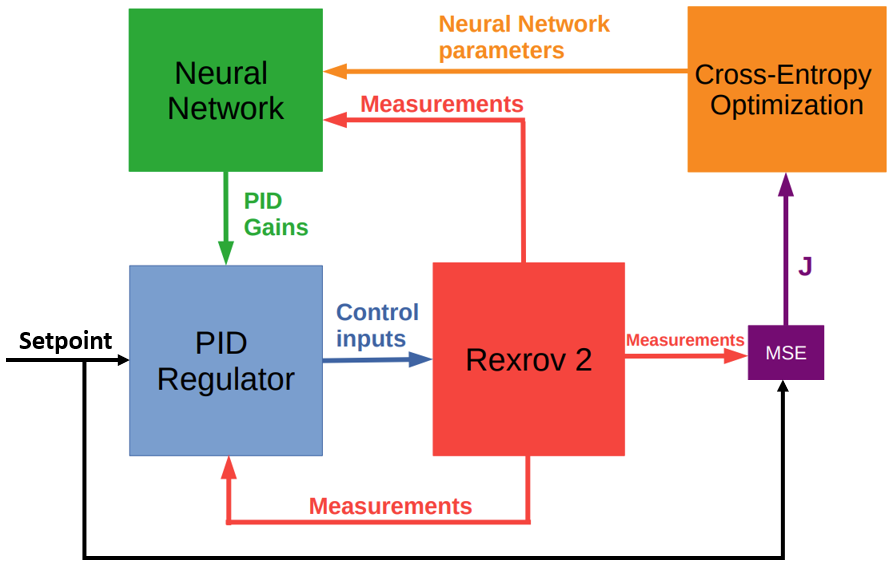}
    \caption{Block diagram of the proposed method for PID tuning loop using Cross-Entropy Deep Learning.}
    \label{fig:control_loop}
\end{figure}

The outputs of the ANNs, are used to compute the PID control law that is ultimately applied to the 6 thrusters. The complete training loop of the ANN is given in Figure \ref{fig:control_loop}.

\subsubsection{Training setup:}We used the following CEM hyperparameters that have been chosen through a grid search: population size $N=25$, proportion to keep $\rho = 0.2$ and added noise $\sigma^2_{noise}=0.1$. Each training episode is composed of 200 timesteps. One epoch is defined as performing one episode using each of the $N$ sets of parameters candidate. The training consists in performing 200 epochs, thus a total of $200\times200\times25=10^6$ timesteps.

\subsubsection{Stability metrics:}Thanks to the proposed Lyapunov function \eqref{Lyapunov-fct} and mapping \eqref{param-space}, we can directly assess the state and parameters stability. We propose to measure the controller's stability as the respective percentage of steps satisfying the state and parameters constraints.

\subsubsection{Evaluation setup:}We propose three evaluation scenarios based on induced disturbance: none, Gaussian noise in sensor measurements, and Gaussian noise in control inputs with current disturbances. The length of the episode is now increased to $2000$ timesteps.

The initial state of the UUV is changed at the beginning of each episode. We compare our LB adaptive controller to its Lyapunov-based counterpart which consist in a PID controller whose parameters are set to: $M_1, M_2, M_3 = diag(0.5 - 10^{-5}, 0.5 - 10^{-5}, 0.5 - 10^{-5}, 0.5 - 10^{-5}, 0.5 - 10^{-5}, 0.5 - 10^{-5}) + 10^{-5} \times \boldsymbol{J}_{6\times6}$ (which satisfies the parameters stability constraints in Eq.\eqref{sol}). The resulting controller remains adaptive (as the gains are a function of the vehicle state \textbf{x}) and will be denoted as naive PID henceforth.

The only difference between these controllers is the value of the parameters used to derive the PID law \eqref{PID}, making the comparison fair.

\subsection{Control performance}
The control performance is measured as the MSE on the setpoint. The evolution of this performance is illustrated in the following figures:
when facing no disturbance in figure \ref{eval_no}; when facing Gaussian noise on the vehicle position and orientation feedback in figure \ref{eval_noise} and when facing Gaussian noise on the control inputs and sea current disturbance in figure \ref{eval_and_currents}.

\begin{figure}
    \centering
    \includegraphics[width=0.9\linewidth]{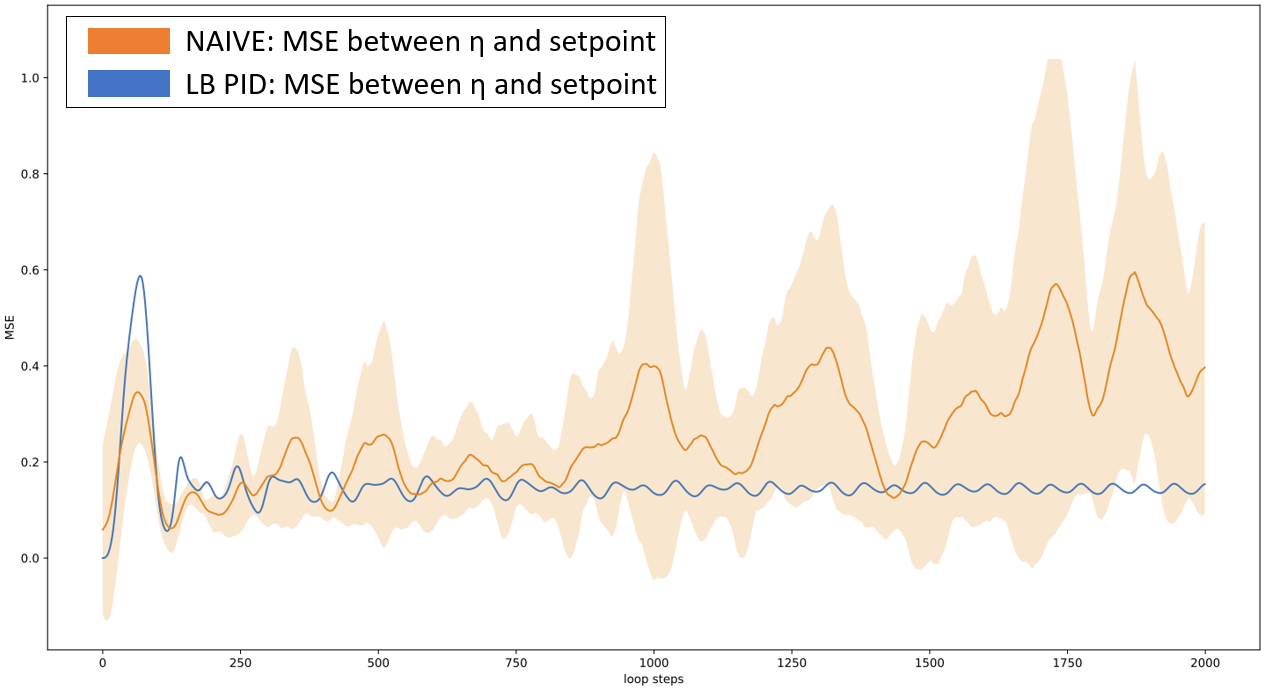}
    \caption{Control performance without disturbance.}
    \label{eval_no}
\end{figure}

\begin{figure}
    \centering
    \includegraphics[width=0.9\linewidth]{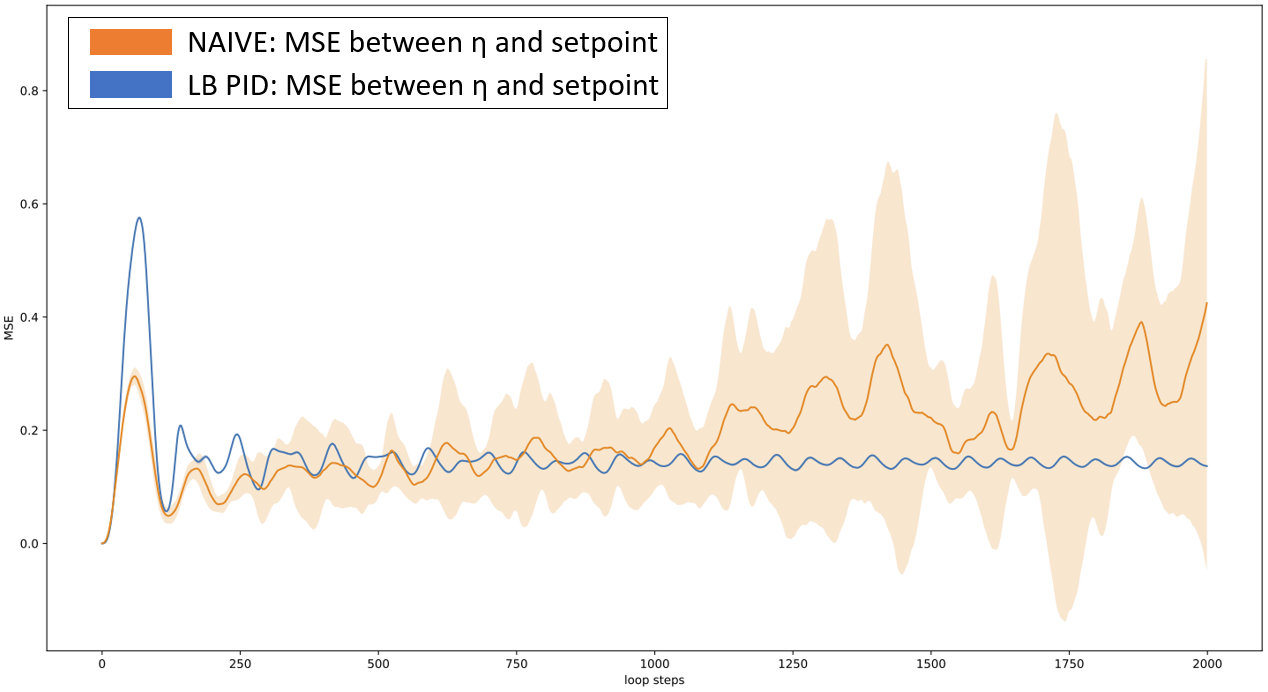}
    \caption{Control performance with noisy position and orientation measurements.}
    \label{eval_noise}
\end{figure}

\begin{figure}
    \centering
    \includegraphics[width=0.9\linewidth]{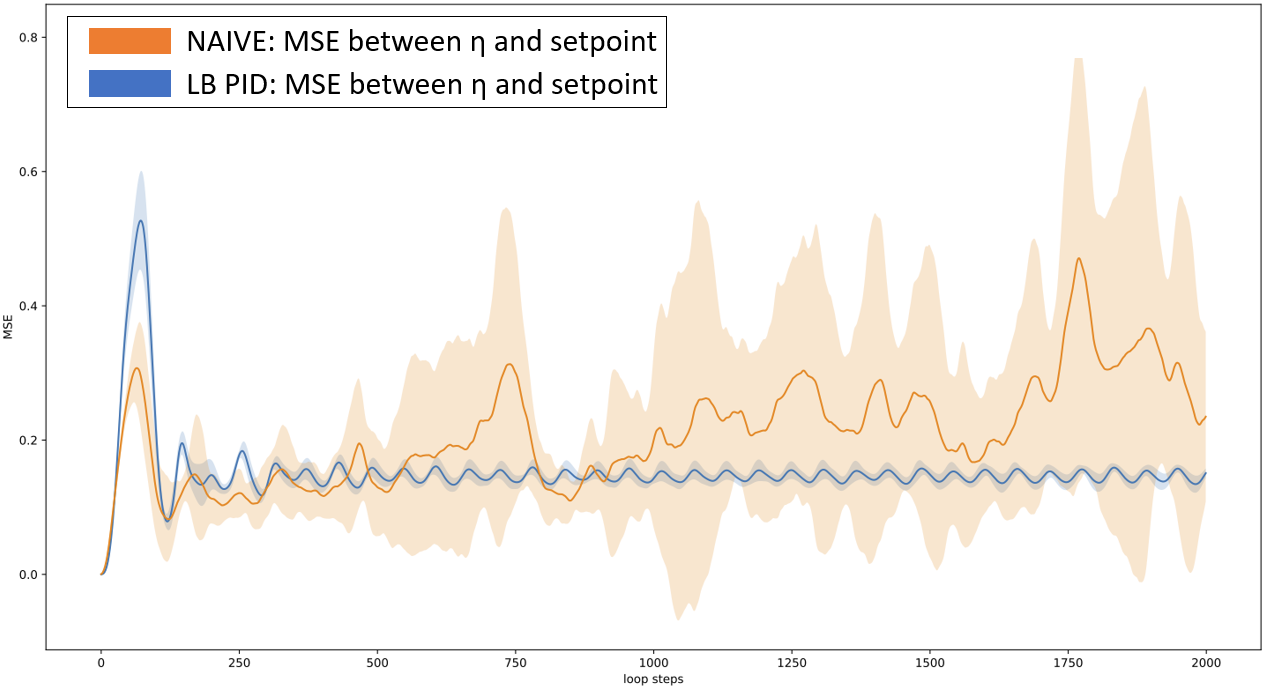}

    \caption{Control performance with noisy control inputs and sea current disturbance.}
    \label{eval_and_currents}
\end{figure}

We can see that the naive PID (in blue) results overall in better performances in terms of setpoint tracking compared to the LB PID (in orange). 
Nevertheless, we can see that the relative performance of the LB PID with respect to the naive PID's performance improves with increased uncertainty. In Figures \ref{eval_noise} and \ref{eval_and_currents}, we can see that the performance of the LB PID matches the naive PID for approximately 500 timesteps, while without disturbance, its performance drops notably earlier as shown in figure \ref{eval_no}.

\begin{figure}
    \centering
    \includegraphics[width=0.9\linewidth]{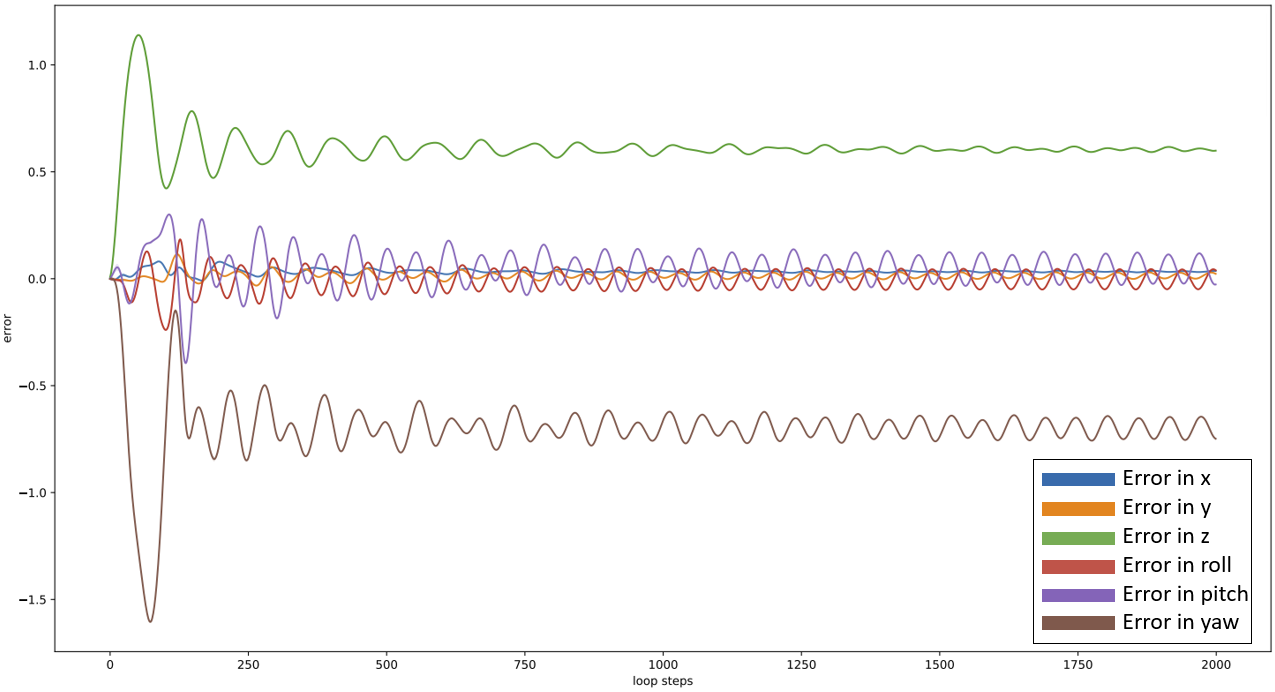} \vspace{-10px}
    \caption{The position and orientation errors of the naive PID during a non-disturbed episode.} 
    \label{fig5:a} 
  \end{figure}

\begin{figure}
    \centering
    \includegraphics[width=0.9\linewidth]{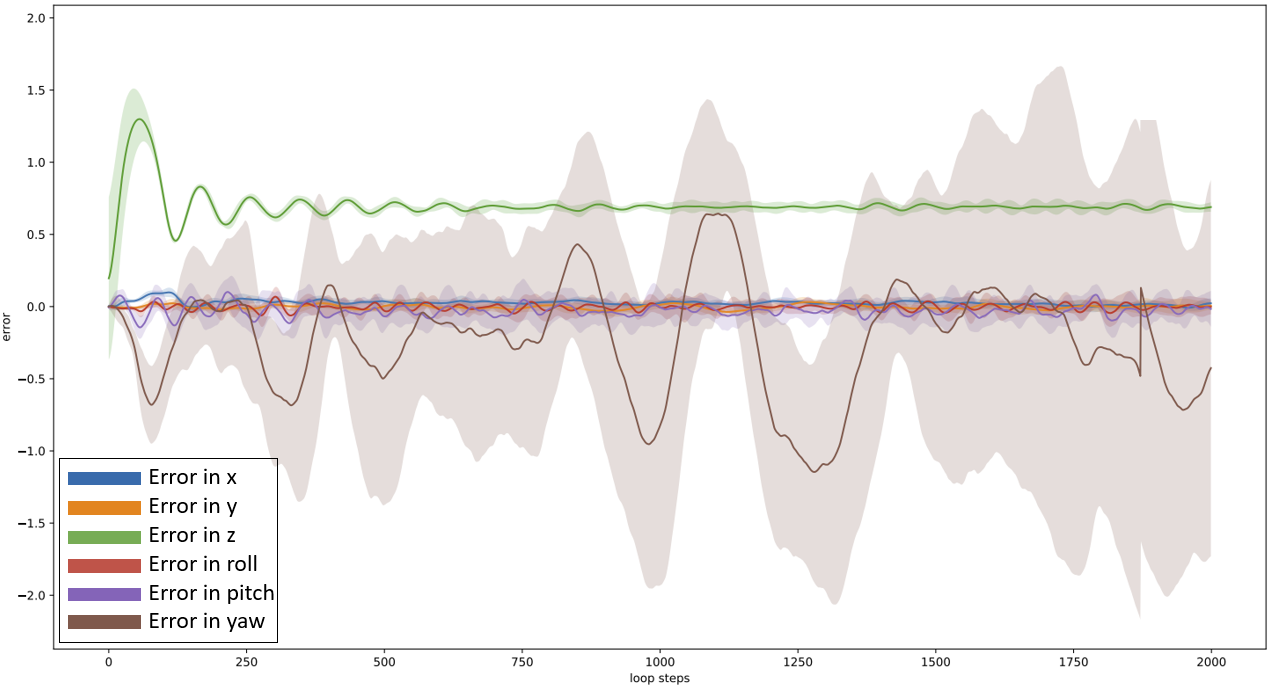}
    \vspace{-10px}
    \caption{The position and orientation errors of the LB PID during a non-disturbed episode.} 
    \label{fig5:b} 
\end{figure}

\begin{figure}
    \centering
    \includegraphics[width=0.9\linewidth]{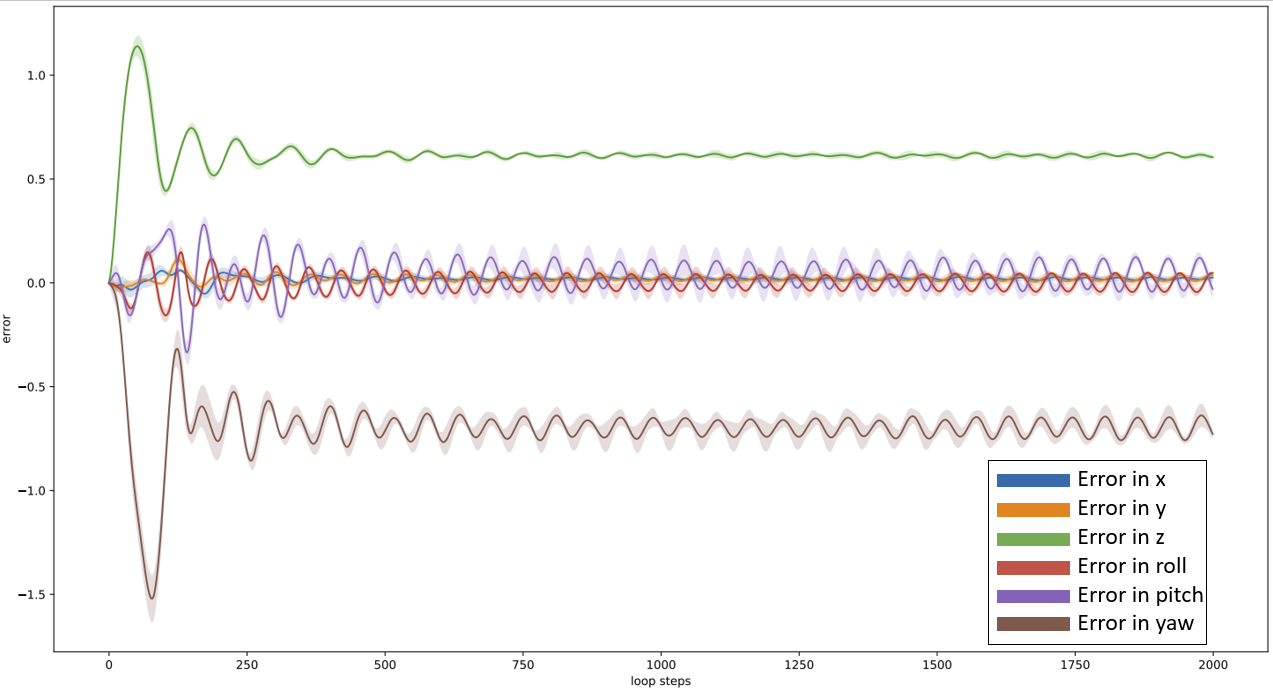} 
    \caption{The position and orientation errors of the naive PID during an episode with measurements noise.}
    \label{fig5:c} 
\end{figure}

\begin{figure}
    \centering
    \includegraphics[width=0.9\linewidth]{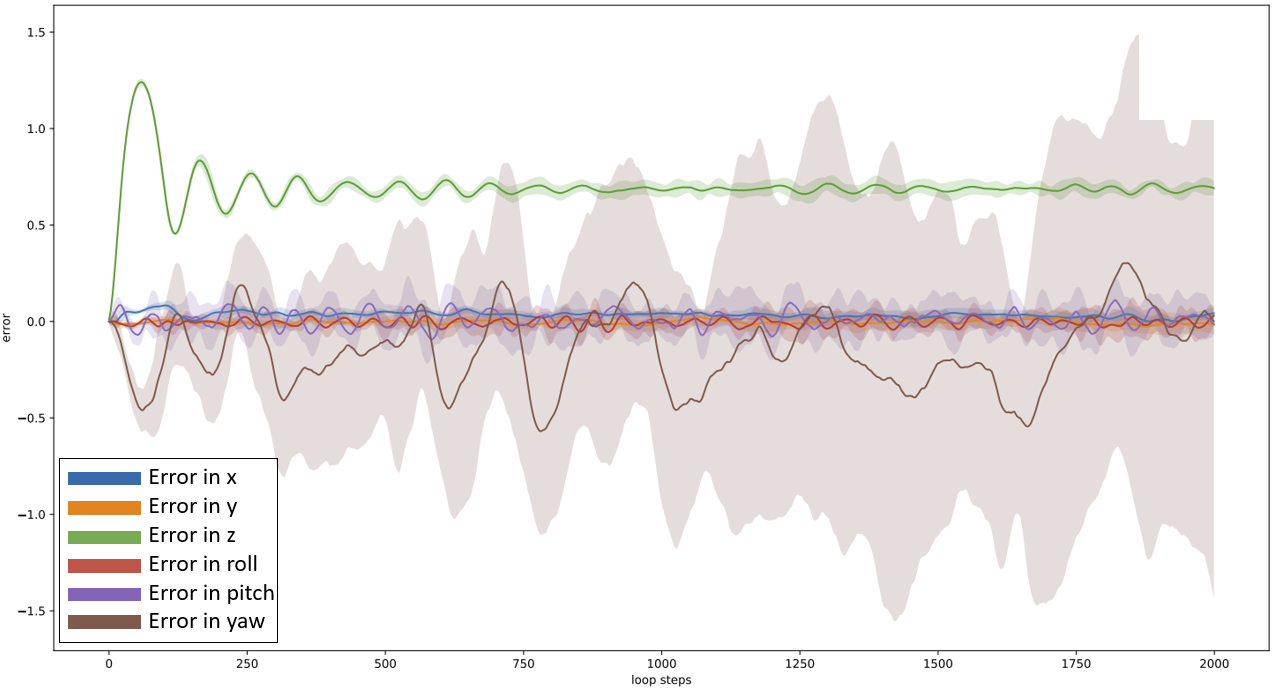} 
    \caption{The position and orientation errors of the LB PID during an episode with measurements noise.}
    \label{fig5:d} 
\end{figure}

\begin{figure}
    \centering
    \includegraphics[width=0.9\linewidth]{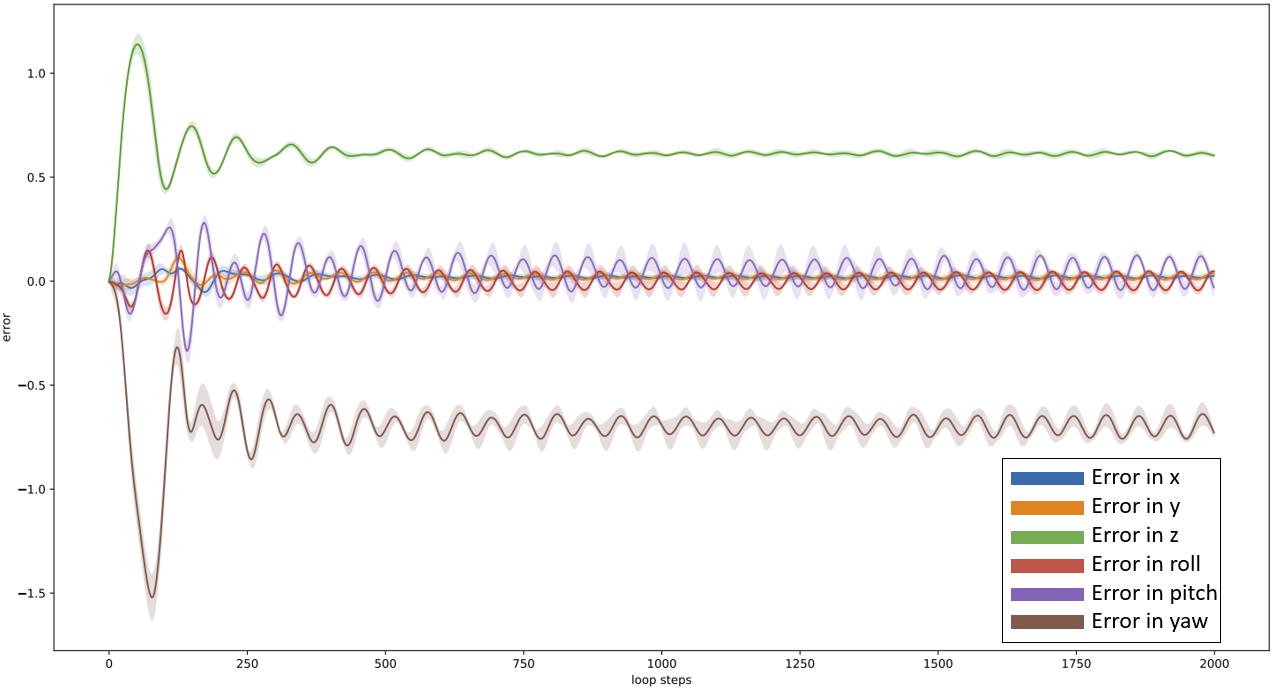} 
    \caption{Position and orientation errors of the naive PID with noisy control inputs and sea current disturbance.}
    \label{fig5:e} 
\end{figure}

\begin{figure}
    \centering
    \includegraphics[width=0.9\linewidth]{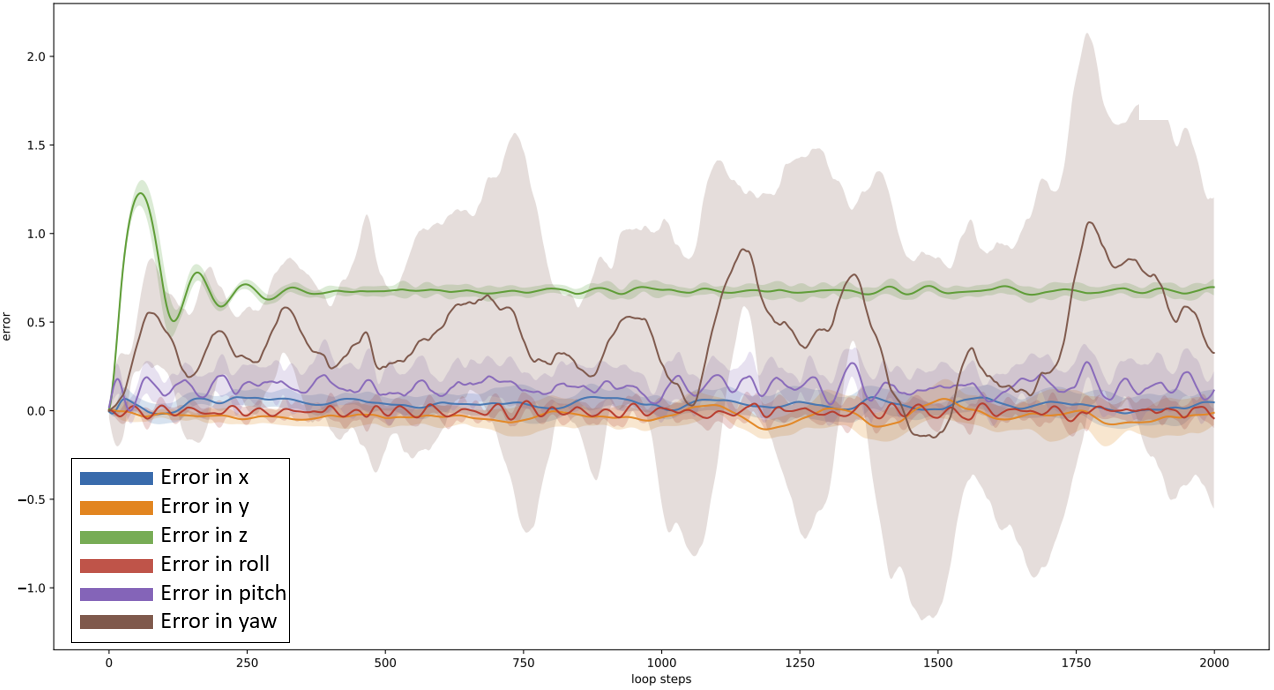}
    \caption{The position and orientation errors of the LB PID with noisy control inputs and sea current disturbance.}
    \label{fig5:f} 
\end{figure}

The evolution of the vehicle's state is represented in Figures \ref{fig5:a}-\ref{fig5:f}. We can observe a difference in regulation dynamics depending on the DoF. For instance, we can see that the depth of the UUV is not successfully regulated by none of the controllers.
Due to the short latency between episodes and the position of the UUV's CoG, the UUV slightly sinks at the beginning of the episode, altering its depth and yaw angle $(z,yaw)$.
This explains the vertical drift in their associated errors observed in the figures. In addition, as seen in Figures \ref{fig5:a}-\ref{fig5:f}, the LB PID tends to regulate effectively more DoF compare to the naive PID (which fails at regulating $(z,yaw)$, see Figures \ref{fig5:a}, \ref{fig5:c}, \ref{fig5:e}).

\subsection{Stability performance}
\begin{figure}
\centering
\includegraphics[width=0.9\linewidth]{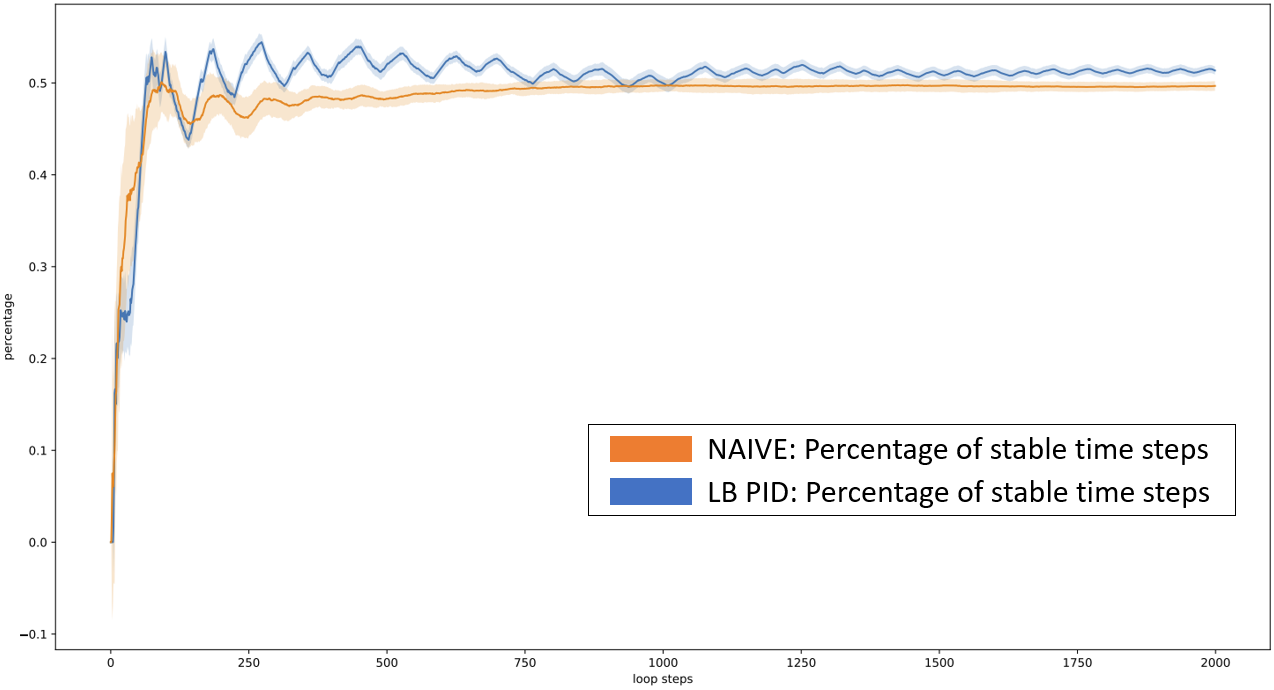}

\caption{Evolution of the state stability during a non-disturbed simulation} 
\label{fig6:a} 
\end{figure}

\begin{figure}
\centering
\includegraphics[width=0.9\linewidth]{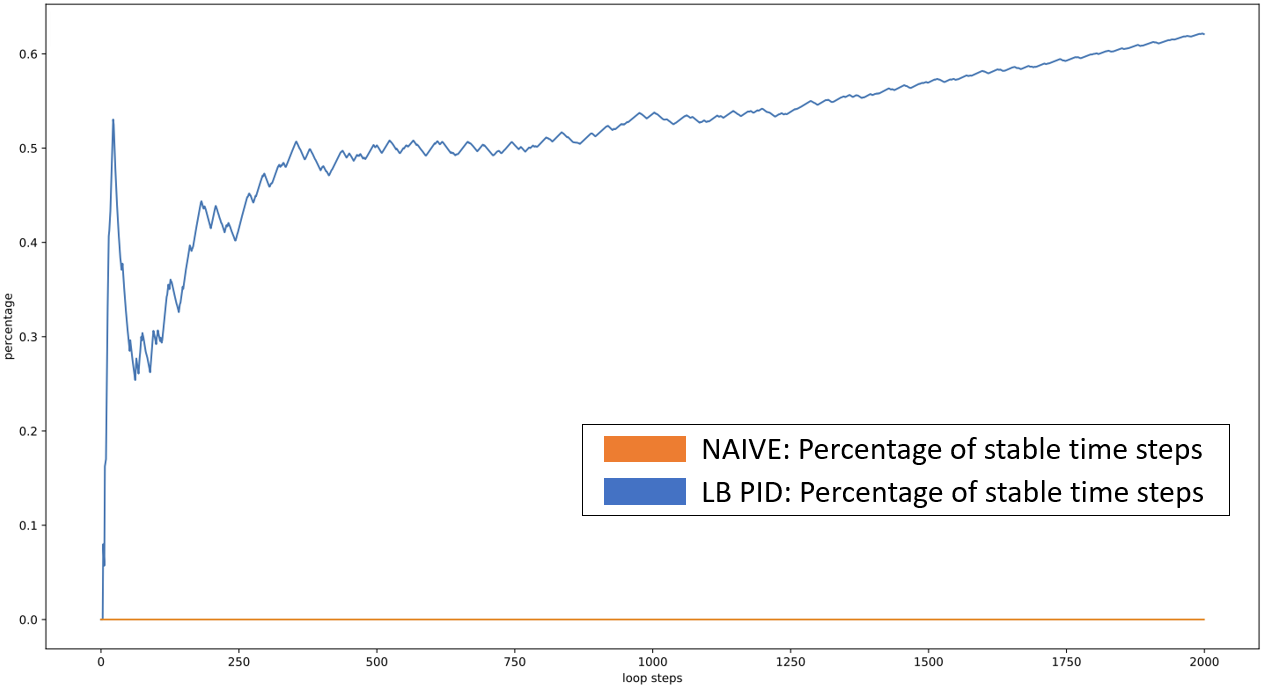}

\caption{Evolution of the parameters stability during a non-disturbed simulation}
\label{fig6:b} 
\end{figure}

\begin{figure}
\centering
\includegraphics[width=0.9\linewidth]{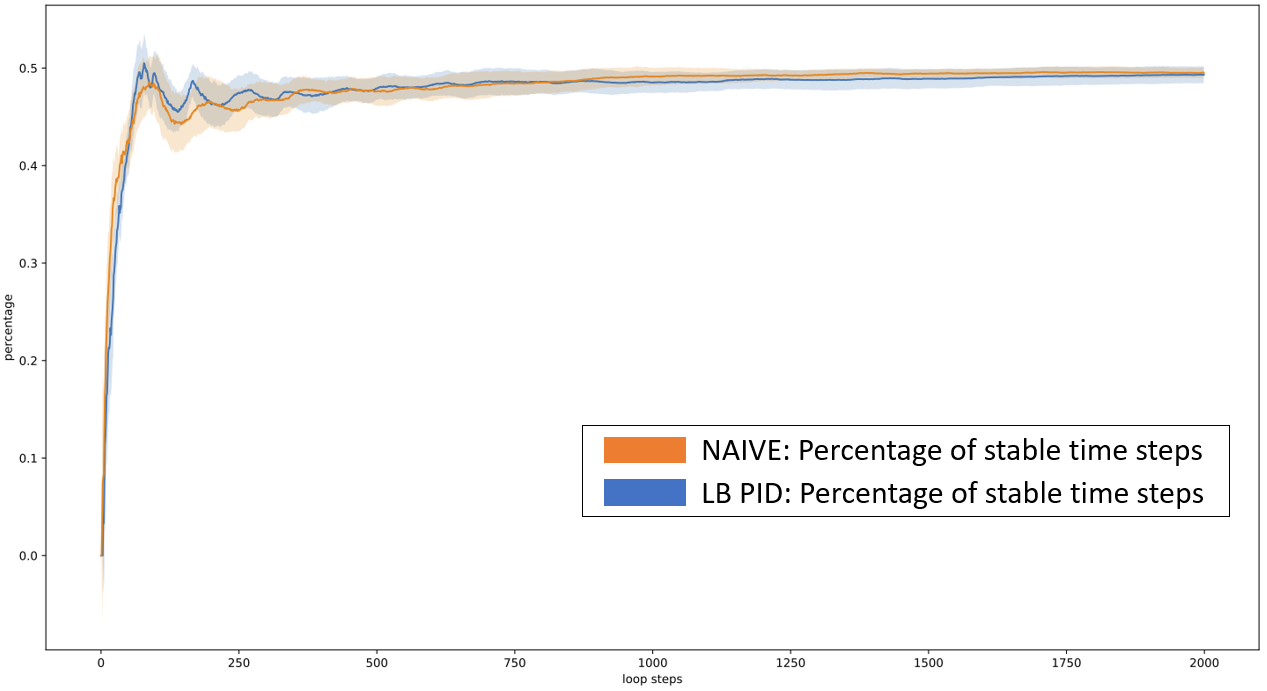}%
\caption{Evolution of the state stability during a simulation with measurements noise}
\label{fig6:c} 
\end{figure}

\begin{figure}
\centering
\includegraphics[width=0.9\linewidth]{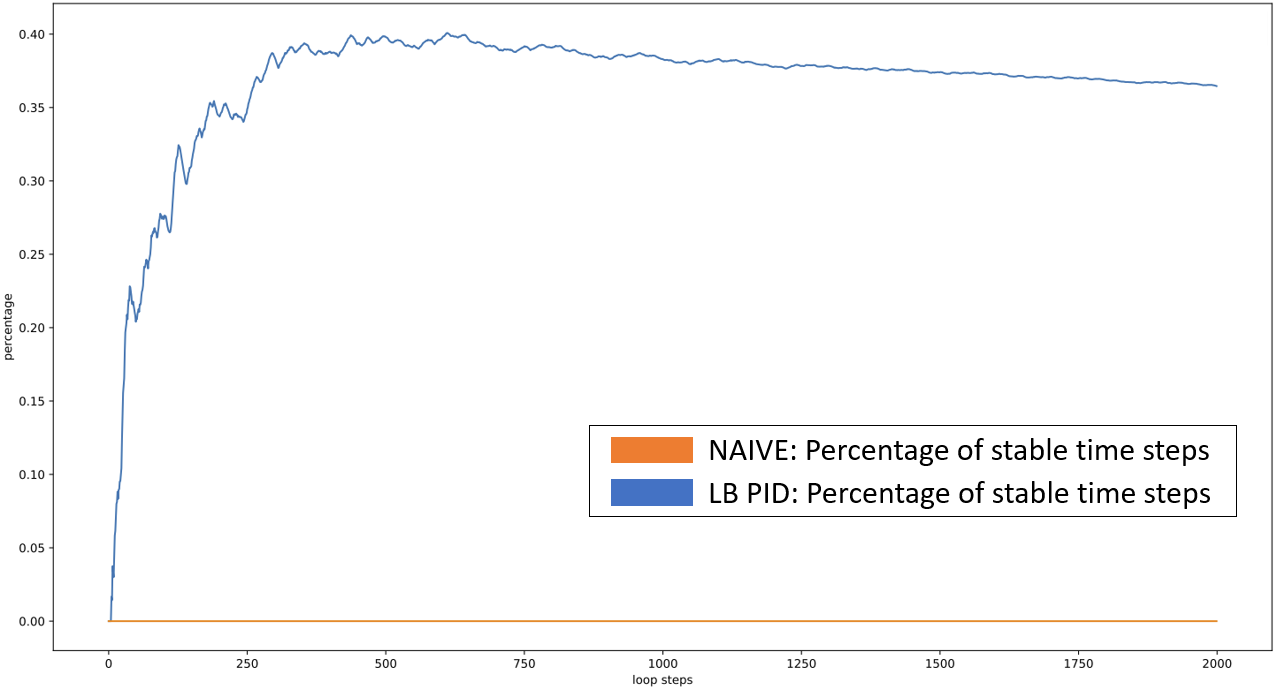}%
\caption{Evolution of the parameters stability during a simulation with measurements noise}
\label{fig6:d} 
\end{figure}

\begin{figure}
\centering
\includegraphics[width=0.9\linewidth]{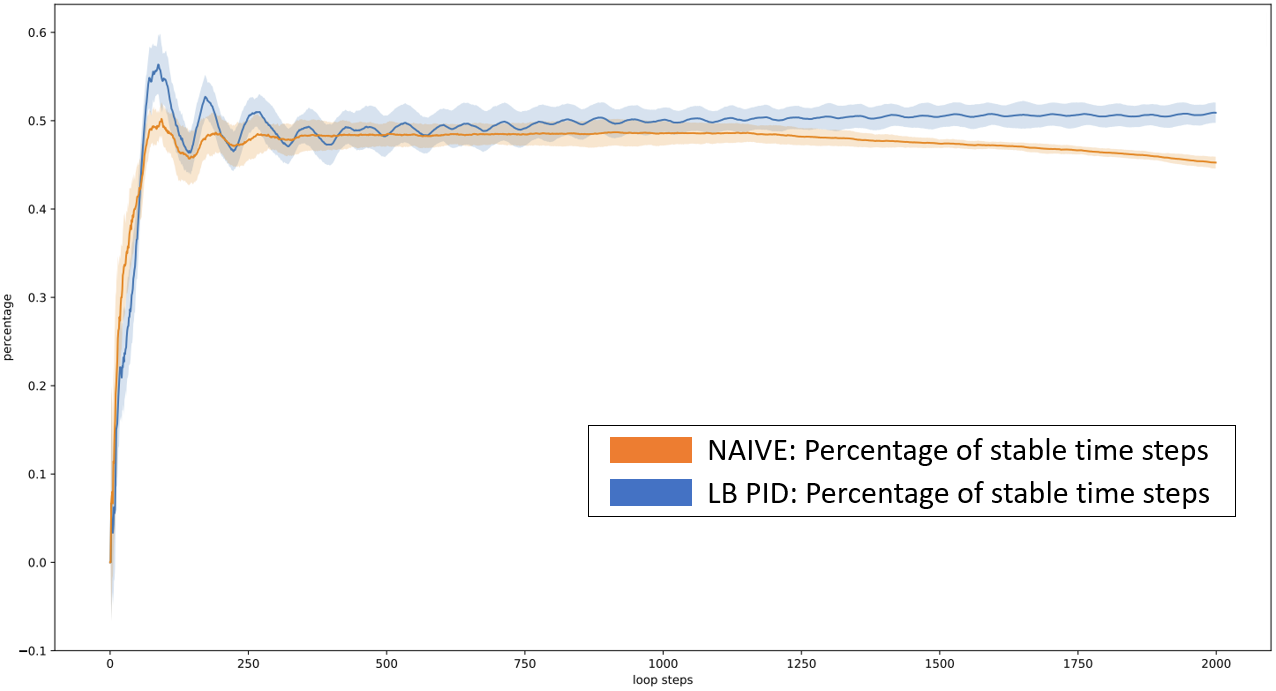}%
\caption{Evolution of the state stability during a simulation with current and thrusters noise}
\label{fig6:e} 
\end{figure}

\begin{figure}
\centering
\includegraphics[width=0.9\linewidth]{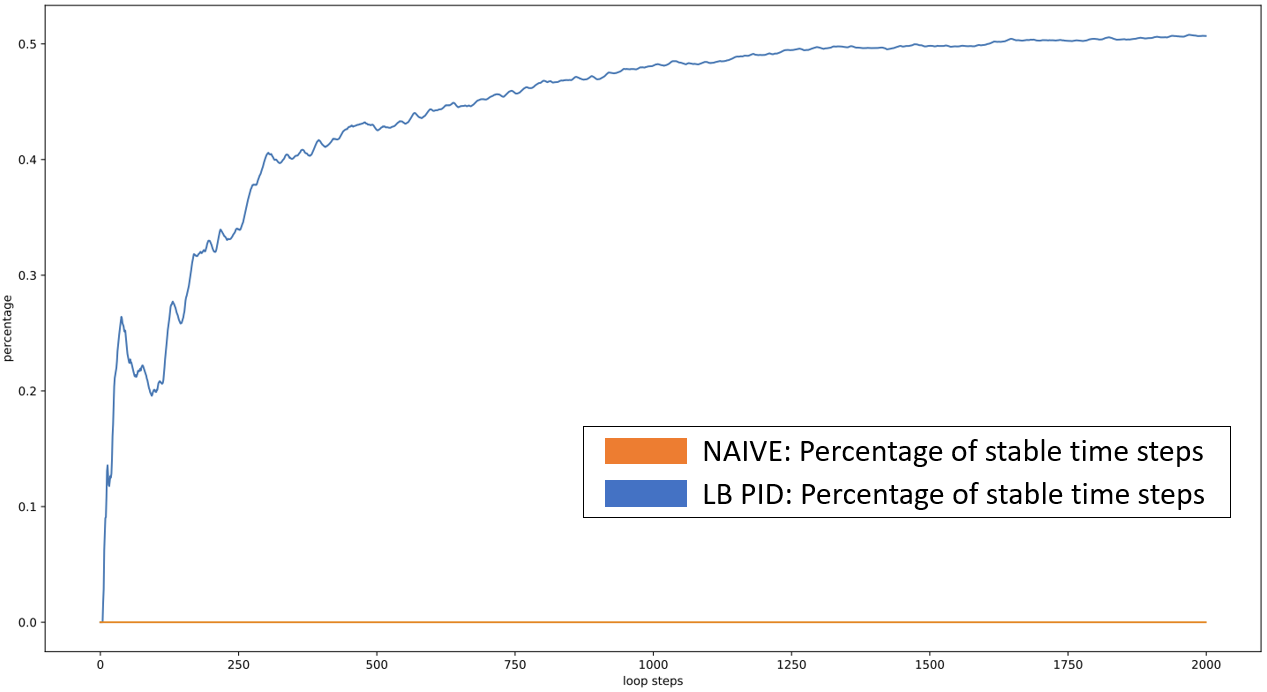}%
\caption{Evolution simulation of the parameters stability during a simulation with current and thrusters noise}
\label{fig6:f} 
\end{figure}

The evolution of the stability metrics is illustrated in Figures \ref{fig6:a}-\ref{fig6:f}. We can see in Figure \ref{fig6:a} that both controllers reach a similar level of state stability. The naive PID exceeds the stability performance of the LB PID in both state and parameters stability. We believe that this is thanks to the stochastic nature of the adaptation (see~Eq.\eqref{action-space}). Note that similar state stability does not guarantee similar control performances. Indeed, as mentioned before, both controllers reach the same level of state stability, however, we have seen that the MSE of the LB PID increases over time, so its control performance decreases but the control loop remains stable. Finally, we can see a mismatch between state and parameters stability: despite reaching a percentage of state stability as good as the naive PID, the gains obtained from the LB PID are not satisfying the constraints \eqref{constraints}. This suggests that the Lyapunov function holds a limited conservatism. In other words, the Lyapunov-based space of parameters seems extremely small compared to the space of parameters obtained from the ANN.

Future work could focus on incorporating the parameter constraints in the learning optimization scheme to analyze the difference in the resulting spaces.

\section{Conclusion}
We proposed the use of CEM for PID tuning to adapt to process variation. Despite not considering stability components in the CEM procedure, the proposed LB PID is able to match the state stability obtained with the Lyapunov-based PID controller. Our result also suggests that the LB PID is better at compensating process variation. Learning-based adaptive control might be a key ingredient toward the safe deployment of fully autonomous UUVs.

\bibliography{cams22_bc}           
\end{document}